\documentclass[lettersize,journal]{IEEEtran}
\usepackage{amsmath,amsfonts}
\usepackage{algorithm}
\usepackage{algpseudocode}
\usepackage{array}
\usepackage[caption=false,font=normalsize,labelfont=sf,textfont=sf]{subfig}
\usepackage{textcomp}
\usepackage{caption}    
\usepackage{stfloats}
\usepackage{url}
\usepackage{verbatim}
\usepackage{graphicx}
\usepackage{booktabs} 
\usepackage{cite}
\usepackage{multirow}
\usepackage{amssymb}  
\usepackage{pifont}
\usepackage[table,xcdraw]{xcolor}
\usepackage{diagbox}
\usepackage{colortbl}
\usepackage{acro}

\begin{document}

\title{MeMo: Attentional Momentum for Real-time Audio-visual Target Speaker Extraction under Impaired Visual Conditions}

\author{Junjie Li$^*$, Wenxuan Wu$^*$, Shuai Wang$^\dagger$, ~\IEEEmembership{Member,~IEEE}, Zexu Pan, Kong Aik Lee,~\IEEEmembership{Senior Member,~IEEE}, \\
Helen Meng,~\IEEEmembership{Fellow,~IEEE}, Haizhou Li$^\dagger$,~\IEEEmembership{Fellow,~IEEE}

\thanks{$^*$: Equal contribution.  $^\dagger$: Corresponding author.}
\thanks{This work was supported in part by National Natural Science Foundation of China
(Grant No. 62401377 and 62271432), Program for Guangdong Introducing
Innovative and Entrepreneurial Teams (Grant No. 2023ZT10X044), Yangtze
River Delta Science and Technology Innovation Community Joint Research
Project (Grant No. 2024CSJGG1100). Partial support was also provided by the Research Grants Council of the Hong Kong SAR (Grant No 15228223), and The Hong Kong Polytechnic University, Project ID P0049192.
 \emph{Junjie Li and Wenxuan Wu contributed equally to this work.} }
\thanks{Junjie Li and Kong Aik Lee are with the Department of Electrical and Electronic Engineering, Faculty of Engineering, The Hong Kong Polytechnic University, Hong Kong SAR, China (email: junjie98.li@connect.polyu.hk, kong-aik.lee@polyu.edu.hk).
}
\thanks{Wenxuan Wu and Helen Meng are with the Department of Systems Engineering and
Engineering Management, The Chinese University of Hong Kong, Hong Kong
SAR, China (email: wwu@se.cuhk.edu.hk; hmmeng@se.cuhk.edu.hk).}

\thanks{Haizhou Li is with  School of Artificial Intelligence (SAI), The Chinese University of Hong Kong, Shenzhen, 518172, China (email: haizhouli@cuhk.edu.cn).}
\thanks{Shuai Wang is with School of Intelligence Science and Technology, Nanjing University, Suzhou, China (email: shuaiwang@nju.edu.cn). }
\thanks{Zexu Pan is with Tongyi Lab, Alibaba Group, Singapore (email: xu325504@hotmail.com).}
}


\maketitle

\begin{abstract}
Audio-visual target speaker extraction (AV-TSE) faces significant performance degradation when visual cues are missing or impaired, limiting its practical deployment in real-world scenarios. This paper introduces MeMo, a novel framework that achieves \textit{attentional momentum} through adaptive memory banks to maintain consistent focus on target speakers despite visual impairments. MeMo employs two complementary memory banks: a speaker bank storing identity features and a contextual bank capturing temporal speech patterns. The framework leverages self-enrolled speech from previous processing windows as supplementary reference, enabling robust extraction even when current visual cues are unreliable. Extensive experiments on the VoxCeleb2 dataset demonstrate MeMo’s effectiveness, yielding approximately 27\% relative improvement over the baseline model under impaired visual conditions in online streaming scenarios.
The contextual bank proves particularly effective, providing over 2 dB gains across multiple evaluation metrics while maintaining real-time processing capabilities. Code and audio demos are available at \url{https://mrjunjieli.github.io/demo_page/MeMo/index.html}.

\end{abstract}

\vspace{-3mm}
\section{Introduction}


\IEEEPARstart{H}{umans} possess  remarkable ability to focus on a specific speaker while filtering out unwanted noise, a phenomenon known as \textit{selective auditory attention} \cite{cherry1953some,bronkhorst2000cocktail}. Beyond auditory information, human attention is inherently multi-modal, with visual inputs enhancing speech intelligibility \cite{schwartz2004seeing,sumby1954visual}. These multi-modal cues are processed interactively in the brain, complementing and enhancing one another to support perception and attention in complex environments.

Inspired by this, audio-visual Target Speaker Extraction (AV-TSE) systems leverage additional visual information, such as dynamic lip sequences, static face images, or body gestures \cite{muse,usev,wu2019time,TSE-review,michelsanti2021overview,PIAVE,co-speech,li23ja_interspeech,chung20c_interspeech,wu2025c,wu2025incorporating,lee2024iianet}, to emulate humans' selective attention capabilities, thereby isolating the target speaker’s speech.
AV-TSE plays a crucial role in enabling downstream tasks such as speech recognition \cite{hu-etal-2023-hearing}, speaker diarization \cite{sell2018diarization}, and speaker verification \cite{sun2023noise}. By incorporating visual information, AV-TSE outperforms audio-only approaches, as visual cues are inherently robust to acoustic interference and provide complementary information to enhance target speech extraction.

Nevertheless, real-world scenarios often present challenges where visual cues may be unavailable or unreliable due to obstructions (e.g., mask wearing), positional constraints (e.g., location outside the camera's field of view), or degraded image quality. Under these conditions, the performance of AV-TSE systems deteriorates significantly \cite{li2024momuse,pan2024ravss,sadeghi2021switching,sadeghi2020robust,wu2022time,ImagineNET,my_lip_concealed,xu2023multi,10596551,liu2023cross,ren2024robust}, limiting their practical utility in real-world applications.

\begin{figure}[tbp]
    \centering
    \includegraphics[width=0.49\textwidth]{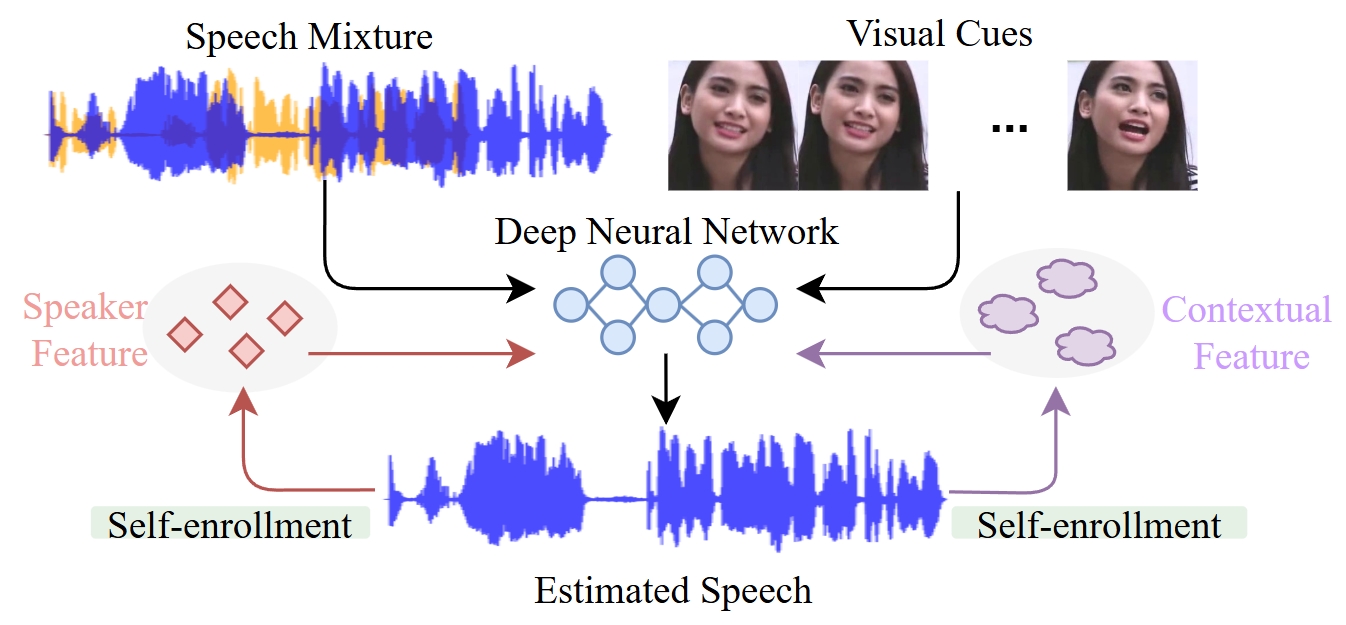}
    \vspace{-6mm}
    \caption{Typical AV-TSE relies solely on visual cues, but performance drops when these cues are unreliable. In contrast, our proposed MeMo enhances reference information by incorporating speaker and contextual features through self-enrollment, ensuring more robust target speaker extraction.}
    \label{fig:self}
    \vspace{-6mm}
\end{figure} 
Neuroscientific research \cite{brungart2007cocktail,kidd2005advantage,best2008object} has shown that humans can enhance their selective auditory attention over time when concentrating on a specific sound source. This improvement is facilitated by the integration of multiple sensory and cognitive cues, including vocal characteristics, visual context, spatial localization, and contextual comprehension \cite{ward2017enhanced,kidd2005advantage,dingemanse2019important}. A fundamental neural mechanism supporting this process is working memory \cite{Brewer19,awh2006interactions}, which plays a crucial role in maintaining and updating the current focus of attention.

Motivated by the human ability to sustain selective auditory attention, we propose \textit{\textbf{MeMo}}, a novel framework that achieves \textit{\textbf{Attentional Momentum}} through the use of \textit{\textbf{Adaptive Memory Banks}}. Attentional momentum refers to the system's capacity to maintain consistent focus on the target speaker over time, even when explicit reference information is unavailable. This ability closely mirrors the human phenomenon of \textit{look once to hear}, where a single glance at visual cue can anchor a sustained auditory focus, enabling enhanced speech extraction despite intermittent visual inputs \cite{look_once_to_hear}. The adaptive memory banks serve as repositories for storing the model’s target attention, enabling continuity in speaker extraction. This functionality allows MeMo to emulate the human-like ability to effectively track a target speaker, even from limited visual information or after just a single observation.

MeMo is not tied to a specific model, instead it represents a generalizable concept that can be integrated into various model architectures. In this paper, we apply MeMo to the real-time single-channel target speaker extraction problem, where only voice characteristics and contextual information are available to compensate for missing visual cues, thereby reinforcing the TSE process. To achieve this, we design two types of memory banks: a \textit{\textbf{speaker bank}} and a \textit{\textbf{contextual bank}}. At the beginning of a conversation, these banks are empty, requiring visual cues to initialize the extraction process. Over time, speaker identity and contextual features are dynamically extracted from historically extracted speech which we refer to as self-enrollment speech,  ensuring continuous adaptation and improved performance, as shown in Fig. \ref{fig:self}.  

The contributions of this paper are summarized as follows: 

\begin{itemize} 
\item We propose the novel concept of attentional momentum for AV-TSE, inspired by human cognitive mechanisms that maintain auditory focus despite visual impairments.
\item  We develop MeMo, an adaptable  framework with dual adaptive memory banks (speaker bank and contextual bank) that implements attentional momentum across various AV-TSE architectures.
\item We validate MeMo's effectiveness through comprehensive experiments on visually impaired VoxCeleb2 datasets, achieving consistent improvements across multiple evaluation metrics. 
\item We further introduce an audio-visual speaker-switching dataset to evaluate robustness in conversations where target speaker changes dynamically, demonstrating MeMo’s adaptability to speaker-switching scenarios.
\end{itemize}

\vspace{-1mm}
\section{Related work}
\subsection{Visual Impairments in AV-TSE}
AV-TSE models typically rely on visual cues to extract target speech, but their performance deteriorates significantly when high-quality visual cues are absent or unreliable \cite{TSE-review}. To address this challenge, Sadeghi and Alameda-Pineda \cite{sadeghi2021switching,sadeghi2020robust} propose bypassing the audio-visual variational auto-encoder (VAE) when visual cues are unreliable and instead using an audio-only VAE model. Wu et al. \cite{wu2022time} and Pan et al. \cite{pan2024ravss} incorporate an attention-based interaction mechanism, which dynamically utilizes adjacent available visual frames to extract target speech when visual cues are partially impaired.
Rather than implicitly inpainting corrupted visual cues, ImagineNET \cite{ImagineNET} explicitly reconstructs them through audio-visual correspondence, employing a repeatedly interlaced structure. Additionally, the VS model \cite{my_lip_concealed} and the audio-visual SpeakerBeam \cite{sato2021multimodal,ochiai2019multimodal} utilize a speaker embedding to complement corrupted visual cues. Beyond using speaker embeddings as auxiliary information, Xu et al. \cite{xu2023multi} further incorporate spatial cues to handle scenarios where visual information is unavailable.
Unlike models that directly integrate visual cues as inputs, Liu et al. \cite{10596551,liu2023cross} propose learning audio-visual correlations using a consistency loss, where visual cues are used only during training.

While the studies mentioned above primarily focus on offline scenarios, we aim to tackle the challenge of missing visual cues in real-time streaming environments. To address this, we introduce a novel approach that mimics the attentional momentum mechanism in the human brain, leveraging various types of information, such as audio, visual and contextual cues, as reference signals to maintain focus on the target speaker.
Our previous work, MoMuSE \cite{li2024momuse}, was an initial attempt to realize this concept, but it is a specific model which can not be generalized to more extraction models.

\vspace{-2mm}
\subsection{Adaptive Memory Bank}
Due to the limited receptive fields of traditional models, they struggle to effectively leverage long-term memory components. To overcome this limitation, memory networks have been proposed \cite{Weston2014MemoryN,sukhbaatar2015end}, inspired by the storage mechanisms found in modern computers. These models are specifically designed to learn how to store, update, and retrieve new features within a memory component, enabling them to handle long-term dependencies and enhance performance.

Wu et al. \cite{wu2018unsupervised} use a memory bank to store visual feature embeddings for all training samples, avoiding repeated computation of representation. By maintaining a global view of instance features, the memory bank enables efficient scaling to large datasets. In multi-modal tasks, memory banks store audio-visual mappings, enabling retrieval of clean speech from visual cues during inference. This method addresses the limitation of insufficient lip movement information in visual speech recognition \cite{CroMM-VSR,MT-LAM,Kim2021MultimodalityAB,yeo2024akvsr,Kim2022DistinguishingHU}, supports speech reconstruction from silent video \cite{hong2021speech,Kim2021MultimodalityAB}, and improves audio-visual speech recognition under noisy conditions \cite{hu-etal-2023-hearing}.
Xu et al. \cite{Xu2018ModelingAA} modeled attention and memory mechanisms for auditory selection using a pre-trained memory bank with fixed slots, similar to the approach in \cite{CroMM-VSR}. Bellur et al. \cite{Bellur2023ExplicitmemoryMA} introduced a unified framework for speech and music source separation, utilizing the concept of temporal coherence to simulate selective auditory attention, gated by embeddings stored in memory.

Incorporating memory banks into these frameworks helps bridge the gap between limited temporal information and the ability to maintain long-term focus on relevant auditory and visual cues, aligning with the attentional momentum mechanism observed in human cognition.

\vspace{-2mm}
\subsection{Self-enrollment Mechanism}
In the self-enrollment mechanism, the model’s output is re-incorporated as new input, enabling the system to continuously refine and update its focus based on past predictions. Unlike methods that rely on irrelevant pre-enrolled speech, this approach has demonstrated greater effectiveness in TSE systems \cite{my_lip_concealed, li2024effectiveness}. Some studies suggest re-encoding the output from earlier layers as input for later ones \cite{zexuSelective, muse}. Other approaches operate in two stages, where the initial output serves as identity information, which is then used to guide the model’s second-stage processing \cite{deng21c_interspeech, my_lip_concealed}. In online real-time scenarios, speech is processed incrementally, with each step utilizing the output from the previous step to continually adapt and refine the extraction process \cite{li2019listening, li2018source, wang2020online, andreev23_interspeech, pan2024neuroheed, pan2024neuroheed+, pan24_interspeech,pan2025online}.

MeMo employs this self-enrollment strategy, extracting both speaker identity and contextual information from the self-enrolled speech. In online settings, the self-enrolled speech provides historical context and crucial speaker identity details, ensuring consistent speaker output.


\section{MeMo}

\subsection{Problem Formulation}

\begin{figure*}[htbp]
\centering
\includegraphics[width=0.98\textwidth]{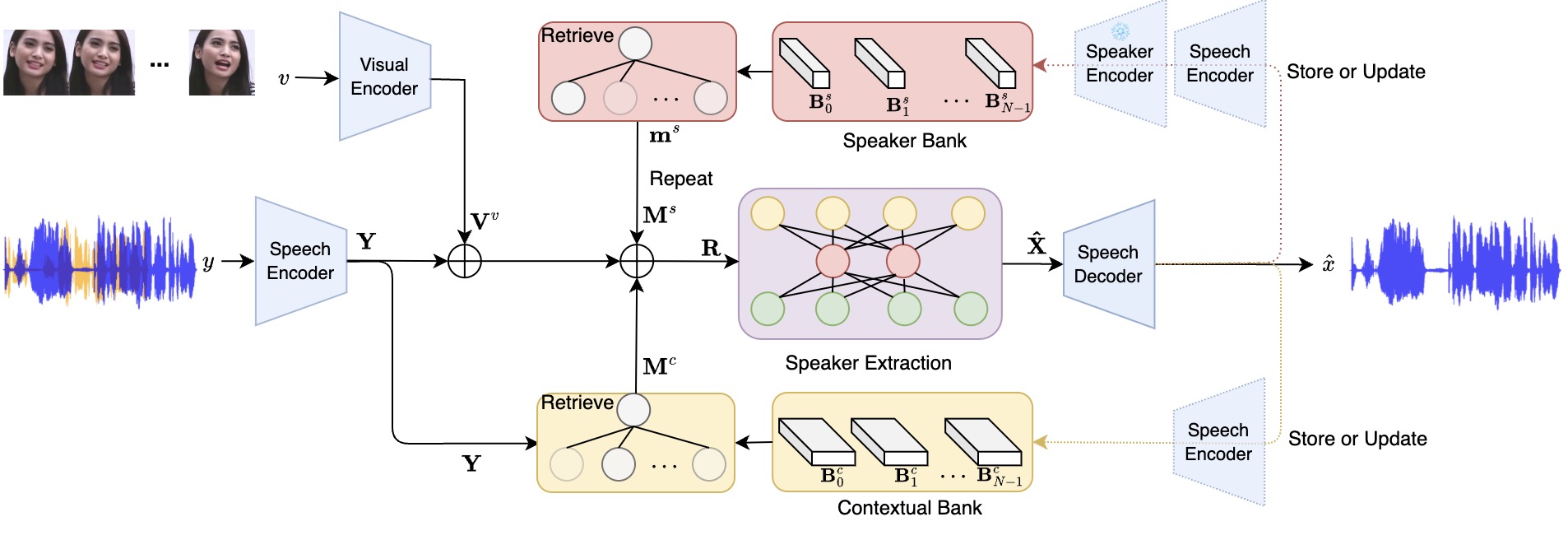}
\vspace{-3mm}
\caption{Overview of MeMo framework. MeMo consists of two types of adaptive memory banks: speaker bank and contextual bank. The speaker bank contains $N$ speaker embeddings, denoted as $\mathbf{B}^s_n$, while the contextual bank contains $N$ speech embeddings, denoted as $\mathbf{B}^c_n$. The operator $\oplus$ denotes concatenation. 
 the parameters of three speech encoders are the same. And the speaker encoder is pre-trained  and frozen during training. }
 \vspace{-6mm}
\label{fig:model}
\end{figure*}

Audio-visual target speaker extraction \cite{TSE-review,michelsanti2021overview} seeks to isolate the target speech from a multi-speaker mixture by leveraging visual cues,  such as the target speaker’s dynamic lip sequences  in this paper. The problem could be formulated as follows:
\begin{equation}
\vspace{-1mm}
    \hat{x} = f_\text{AV-TSE}(y,v ;\theta),
    \vspace{-1mm}
\end{equation}
where $y$ denotes mixture,  $\hat{x}$ represents the extracted speech, $v$ denotes target visual cues, $\theta$ is the parameter of AV-TSE model. 
The mixture $y$ consists of multiple speech signals: 
\begin{equation}
\vspace{-1mm}
    y_{[t]} = x_{[t]} + \sum_{i=1}^I s_{i[t]} ,
    \vspace{-1mm}
\end{equation}
where $x_{[t]}$ and $s_{i[t]}$ denote target speech and $i$-th interference speech signals, respectively, and $t$ is a discrete-time index.  In this paper, we consider several types of visual impairments, including visual missing, lip concealment, and low-resolution inputs, which are introduced in Section~\ref{sec:data}. For all these scenarios, no explicit masking mechanism or confidence-based gating is applied to the visual input $v$.

For a general AV-TSE system, there are four basic modules: the speech encoder, visual encoder, speaker extractor, and speech decoder. 
Let $y \in \mathbb{R}^{T}$ denote a time-domain speech mixture waveform, where $T$ denotes the length of speech. The speech encoder transforms the input signal into a latent speech embedding
$\mathbf{Y} \in \mathbb{R}^{L \times C}$, where $L$ and $C$ denote the temporal length and channel dimension of the latent representation, respectively.
Similarly, the visual encoder processes a visual sequence
$v \in \mathbb{R}^{\frac{T \cdot \mathrm{sr}^v}{\mathrm{sr}^a} \times H \times H}$,
where $\mathrm{sr}^v$ and $\mathrm{sr}^a$ denote the sampling rates of the visual and audio sequences, respectively, into a visual embedding. Due to the difference between visual and audio sampling rates, the resulting visual embedding is temporally up-sampled by linear interpolation  to match the speech temporal resolution $L$, yielding
$\mathbf{V}^v \in \mathbb{R}^{L \times C}$. $H$ denotes the width and height of one visual frame. The speaker extractor estimates a mask to generate the estimated speech embedding $\hat{\mathbf{X}} \in \mathbb{R}^{L \times C}$, with visual cues as reference. 
The speech decoder then reconstructs the estimated speech waveform $\hat{x} \in \mathbb{R}^{T}$.

\vspace{-2mm}
\subsection{System Overview}
In this study, we aim to develop a robust AV-TSE system for real-world scenarios, where extraction operates as a streaming process and target visual cues may be unavailable.

Inspired by human attentional momentum, we propose MeMo, a novel framework that is not tied to a specific model but can be integrated into various separation backbones. MeMo introduces two types of adaptive memory banks: a speaker bank and a contextual bank, as illustrated in Fig. \ref{fig:model}. Unlike conventional AV-TSE systems that rely solely on visual cues as reference, MeMo retrieves speaker identity and contextual information from memory banks as supplementary references. This mechanism effectively addresses challenges posed by impaired visual cues in real-world conditions.

Notably, both memory banks are adaptive rather than pre-trained and frozen. At the start of the online processing, these banks are empty. Once the estimated speech is extracted from the first processing window, it is self-enrolled as a speaker embedding and a contextual speech embedding, which are then stored in their respective banks.  As the memory banks fill up, an updating mechanism dynamically replaces older embeddings with new ones, ensuring continuous adaptation.

As introduced above, attentional momentum is achieved by consistently extracting attention information from the historically estimated speech, ensuring that attention remains focused on the same speaker over time. Additionally, current visual cues are also utilized. The balance between historical and current information makes the model well-suited for real-world conditions, even in target speaker switching scenarios, where the target speaker may change during a conversation. We will discuss this in Section \ref{sec:att_mo}. 


\vspace{-2mm}
\subsection{Adaptive Memory Bank}
\label{sec:memban}
As mentioned in the system overview, we employ two types of adaptive memory banks to store the tracking signals for the target speaker: speaker identity cues and historical contextual information of the conversation. In this section, we will introduce how to retrieve the most relevant speaker cues and contextual cues from these memory banks.

\subsubsection{Speaker Bank}

\begin{figure}[htbp]
\vspace{-2mm}
\centering
\includegraphics[width=0.49\textwidth]{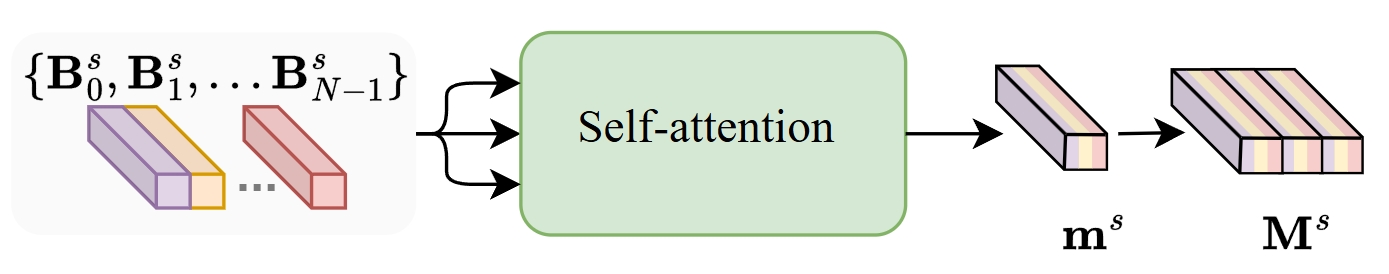}
\vspace{-6mm}
\caption{Illustration of retrieval operation from speaker bank. }
\label{fig:attn_s}
\vspace{-2mm}
\end{figure}
 Suppose the speaker bank has $N$ slots, each slot contains a speaker embedding.  The speaker bank could be represented as:   
        $\mathbf{B}^s = \{\mathbf{B}^s_0, \mathbf{B}^s_1, \dots, \mathbf{B}^s_{N-1}\}$, where $\mathbf{B}^s_n \in \mathbb{R}^{ C}$ denotes the $n$-th slot and $\mathbf{B}^s \in \mathbb{R}^{N \times C}$ denotes the whole speaker bank, as shown in Fig. \ref{fig:attn_s}. 


        Considering the intra-speaker variability of different speaker slots~\cite{TSE-review}, we want to effectively capture the most relevant speaker cues correlated with the target speaker in the current mixture frames, we apply a self-attention \cite{vaswani2017attention} mechanism over the entire speaker bank $\mathbf{B}^s$.  Specifically, we compute the self-attention using the speaker bank as the query, key, and value:
  \setlength{\jot}{1pt}
  \begin{align}
  \mathbf{Q}^s \in \mathbb{R}^{N\times C} &= \mathbf{B}^s\mathbf{W}_Q^s, \\
  \mathbf{K}^s  \in \mathbb{R}^{N\times C} &= \mathbf{B}^s\mathbf{W}_K^s, \\
  \mathbf{V}^s  \in \mathbb{R}^{N\times C} &= \mathbf{B}^s\mathbf{W}_V^s,
  \end{align}
  where $\mathbf{W}_Q^s,\mathbf{W}_K^s,\mathbf{W}_V^s \in \mathbb{R}^{C\times C}$ denote the weight matrices. 
  Then, a softmax operation is applied along the first dimension, and the attention score matrix among different speaker slots is computed as follows:
  \vspace{-3mm}
\begin{equation}  
\mathbf{A}^s \in \mathbb{R}^{N \times N}  = softmax\left( \frac{\mathbf{Q}^s{\mathbf{K}^s}^\top}{\sqrt{C}} , \text{dim=1} \right). 
\vspace{-1mm}
\end{equation}
Then we compute the average score on  dimension zero to get the final attention weight for each speaker embedding: 
\vspace{-1mm}
\begin{equation}
    \overline{\mathbf{A}^s} \in \mathbb{R}^{1\times N}= AvgPooling(\mathbf{A}^s, \text{dim=0}). 
    \label{equ:att_s}
    \vspace{-1mm}
\end{equation}
Then the most relevant speaker cue $\mathbf{m}^s$  is obtained: 
\vspace{-1mm}
\begin{equation}
    \mathbf{m}^s \in \mathbb{R}^{1\times C} = \overline{\mathbf{A}^s}\mathbf{V}^s.
    \vspace{-1mm}
\end{equation}
To align with mixture embedding $\mathbf{Y}$, $\mathbf{m}^s$ is repeated along length dimension to get $\mathbf{M}^s \in \mathbb{R}^{L \times C}$.

\begin{figure}[htbp]
\centering
\vspace{-2mm}
\includegraphics[width=0.49\textwidth]{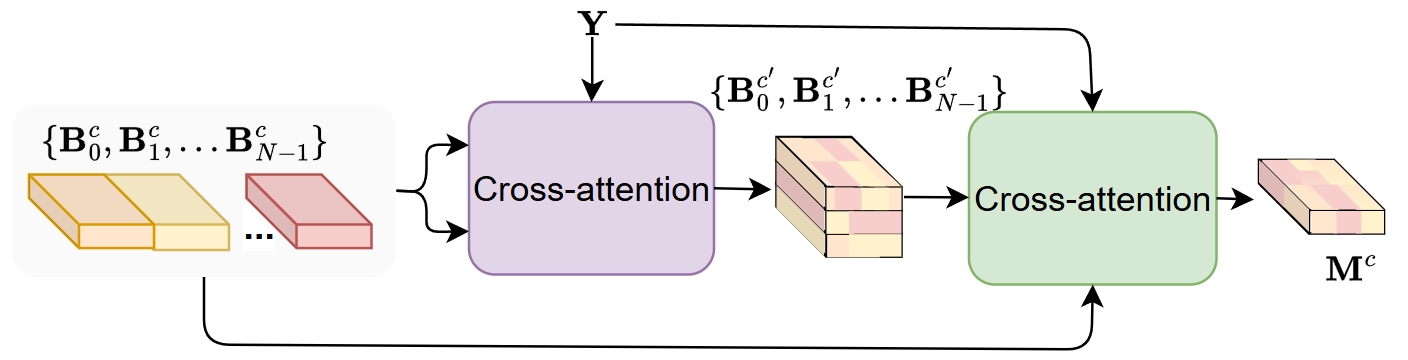}
\caption{Illustration of retrieval operation from contextual bank.  }
\label{fig:attn_c}
\vspace{-2mm}
\end{figure}

\subsubsection{Contextual Bank}
Similar to the speaker bank,  the contextual bank contains $N$ slots:
$\mathbf{B}^c = \{\mathbf{B}^c_0, \mathbf{B}^c_1, \dots, \mathbf{B}^c_{N-1}\}$, where $\mathbf{B}^c_n \in \mathbb{R}^{L \times C}$ and $\mathbf{B}^c \in \mathbb{R}^{L \times N \times C}$ denote the  $n$-th slot and whole contextual bank, respectively. In this case, we employ two cross-attention layers to retrieve the most relevant contextual information, as shown in Fig. \ref{fig:attn_c}. 

The first cross-attention layer is employed for $N$ times to capture the most relevant content from each  contextual slot $\mathbf{B}^c_n$ based on the mixture embedding $\mathbf{Y} \in \mathbb{R}^{L \times C}$:
\setlength{\jot}{1pt}
\begin{align}
    \mathbf{K}^{c_1} \in \mathbb{R}^{L\times C} &= \mathbf{Y} \mathbf{W}^{c_1}_K, \\ 
    \mathbf{Q}^{c_1}_n  \in \mathbb{R}^{L\times C} &=  \mathbf{B}_n^c \mathbf{W}^{c_1}_Q, \\
    \mathbf{V}^{c_1}_n  \in \mathbb{R}^{L\times C} &=\mathbf{B}_n^c \mathbf{W}^{c_1}_V,
\end{align}
where $\mathbf{W}^{c_1}_K, \mathbf{W}^{c_1}_Q, \mathbf{W}^{c_1}_V \in \mathbb{R}^{C \times C}$ are weight matrices. 
Then $n$-th filtered contextual speech embedding is obtained:
\begin{equation}
    \mathbf{B}^{c'}_n \in \mathbb{R}^{L \times C}  = softmax\left( \frac{\mathbf{Q}^{c_1}_n{\mathbf{K}^{c_1}}^\top}{\sqrt{C}}, \text{dim=1} \right) \mathbf{V}_n^{c_1}. 
\end{equation}
The second attention is also a cross-attention, aiming to obtain global contextual information from different slots. In this case, we take mixture speech embedding $\mathbf{Y} \in \mathbb{R}^{L \times C}$, contextual speech embedding $\mathbf{B}^c \in \mathbb{R}^{L \times N \times C}$ and filtered contextual speech embedding $\mathbf{B}^{c'}  \in \mathbb{R}^{L \times N \times C}$ to compute key, value and query, respectively: 
\setlength{\jot}{1pt}
\begin{align}
    \mathbf{K}^{c_2} \in \mathbb{R}^{L \times C} &= \mathbf{Y} \mathbf{W}^{c_2}_K,  \\
    \mathbf{Q}^{c_2} \in \mathbb{R}^{ L \times N\times C} & = \mathbf{B}^{c'}\mathbf{W}^{c_2}_Q,\\
    \mathbf{V}^{c_2} \in \mathbb{R}^{L \times N\times C} & =\mathbf{B}^{c}\mathbf{W}^{c_2}_V. 
\end{align}
Then the softmax operation is used to compute the weight for each slot \footnote{We omit some dimension-transposition operations for simplicity.}:
\vspace{-3mm}
\begin{equation}
    \mathbf{A^c}\in \mathbb{R}^{L \times 1 \times N } = softmax\left( \frac{\mathbf{Q}^{c_2}\mathbf{K}^{c_2}}{\sqrt{C}}, \text{dim=1} \right). 
    \label{equ:att_c}
\end{equation}
Here, the attention score $\mathbf{A}^c$ denotes the weight for each contextual slot. Then the most relevant contextual speech embedding is obtained by weighting different slots: 
\begin{equation}
    \mathbf{M}^c \in \mathbb{R}^{L \times C}= \mathbf{A}^c \mathbf{V}^{c_2}.
    \vspace{-5mm}
\end{equation}



\vspace{-2mm}
\subsection{Attentional Momentum}
\label{sec:att_mo}
The attentional momentum mechanism refers to the human ability to maintain focus on a specific speaker over time during a conversation. Our goal is to replicate this ability in AV-TSE systems. We first implemented this concept in MoMuSE \cite{li2024momuse}, achieving convincing results.

\begin{figure}[htbp]
\vspace{-3mm}
    \centering
    \includegraphics[width=0.49\textwidth]{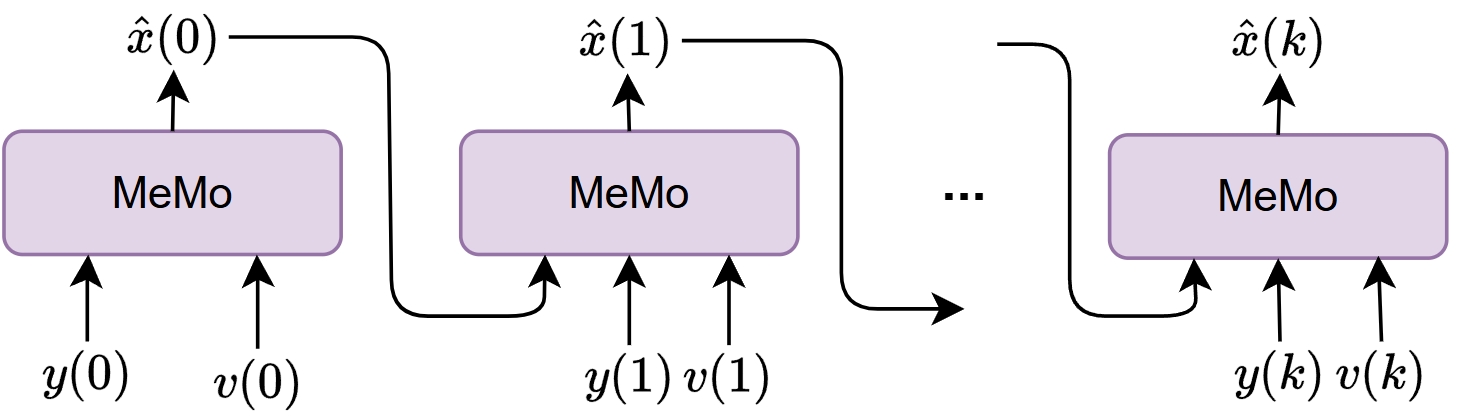}
    \vspace{-5mm}
    \caption{The basic idea of attentional momentum. $y(k)$, $v(k)$ and $\hat{x}(k)$ denote the speech mixture, visual cues and estimated speech at the $k$-th online window step. }
    \vspace{-3mm}
    \label{fig:mo}
\end{figure}

In online scenarios, as shown in previous work \cite{li2024momuse, pan2024neuroheed, pan2024neuroheed+}, input speech is processed using a sliding window. This window shifts forward gradually over time, with each step incorporating new data for speaker extraction. Fig. \ref{fig:mo} shows the core concept of attentional momentum. At the start of online processing, MeMo initially relies on visual cues $v(0)$ for speaker extraction. As the process continues, the model incorporates previously estimated speech as additional reference, helping to maintain a consistent focus on the target speaker.

Based on the adaptive memory banks, the attentional momentum mechanism could be summarized into three components: storing, retrieval, and updating:
\vspace{-1mm}
\begin{itemize}
    \item \textbf{Storing}: 
    Suppose the volume of each memory bank is $N$.  Then during the $k$-th ($  N-1\geq k \ \geq 0$) online processing window step, these memory banks are not full.  The estimated speech $\hat{x}(k)$ at $k$-th window step, is processed by the speech encoder and speaker encoder to generate a contextual speech embedding $\mathbf{B}_k^c$  and speaker embedding  $\mathbf{B}_k^s$,  as illustrated by the dotted lines in Fig. \ref{fig:model}.  These embeddings are then stored in their respective memory banks.
    We utilize a pre-trained ECAPA-TDNN \cite{desplanques2020ecapa} from WeSpeaker \cite{wang2023wespeaker} as the speaker encoder \footnote{\url{https://wenet.org.cn/downloads?models=wespeaker&version=voxceleb_ECAPA512.zip}}.  The structure and parameters of the speech encoder in Fig. \ref{fig:model} are totally the same. 
    It is important to note that $\hat{x}(0)$ is predicted solely using visual cues $v(0)$, cause the memory banks are empty at that time. After the $0$-th window step,  the extraction processing is guided by multiple cues, including visual cues, speaker cues and contextual cues retrieved from the adaptive memory banks. 

\item \textbf{Retrieval}:  
As introduced in Section \ref{sec:memban}, MeMo employs an attention-based mechanism to retrieve the most relevant information from all slots in the memory banks. Instead of selecting a single slot, MeMo leverages attention weights to fuse information from all slots, enhancing the model’s ability to maintain long-term consistency.
At the $k$-th window step, where $k > 0$, memory banks are not empty. MeMo generates momentum speaker features $\mathbf{M}^s(k)$ and momentum contextual features  $\mathbf{M}^c(k)$ to keep model's attention at $k$-th  window step in line with previous steps. Hence, the final concatenated features are shown below: 
\vspace{-2mm}
\begin{equation}
    \mathbf{R}(k) = \mathbf{Y}(k) \oplus \mathbf{V}^v(k) \oplus \mathbf{M}^s(k) \oplus \mathbf{M}^c(k),
    \vspace{-2mm}
\end{equation}
where $\oplus$ denotes concatenation. Embeddings $\mathbf{Y}$, $\mathbf{V}^v$, $\mathbf{M}^s$ and $\mathbf{M}^c$ denote mixture feature, visual feature, speaker feature and contextual features, respectively. 


\item \textbf{Updating}:  
Suppose the memory bank can store up to $N$ feature slots. At the $k$-th window step, where $k\geq N$, these banks are full. 
We design two methods to pop out the old embedding and insert the new one. 
\begin{itemize}
    \item FIFO: Like a queue in a computer system, once the memory bank reaches its maximum capacity, the oldest feature is discarded following a first-in-first-out policy. This ensures that the memory bank always contains the most recent features for extraction.
    \item ABS: To further enhance adaptability, we propose an alternative updating method called attention-based selection (ABS). In this method, instead of always discarding the oldest feature, the feature with the lowest attention score (computed as shown in Equation \ref{equ:att_c} and Equation \ref{equ:att_s}) is removed from the bank.
\end{itemize}
\end{itemize}

\vspace{-2mm}
\subsection{Training and Inference Strategy}
MeMo is designed for online processing scenarios, making it highly applicable to real-world applications. The attentional momentum mechanism enables MeMo to maintain attention on the same speaker over time. To achieve this, autoregressive (AR) training strategy is a straightforward approach. However, the computational cost is high, and errors can accumulate easily. To address this, we adopt a pseudo-autoregressive (PAR) training approach, similar to PARIS \cite{pan24_interspeech}.

\begin{algorithm}[htbp]
    \caption{PAR Training Strategy for MeMo}
    \label{alg} 
    \textbf{Input:} Speech mixture $y$, visual inputs $v$, pre-enrolled speech $p$ (optional), ground truth $x$, model parameters $\theta$
    
    \textbf{Output:} Trained model parameters $\theta^*$
    
    \begin{algorithmic}[1]
    \State \textbf{Stage 1: Initial Extraction}
    \If{VP$_\text{Init}$ setting}
        \State $\hat{x}^1 \leftarrow f_\text{AV-TSE}(y, v, p; \theta)$
    \Else
        \State $\hat{x}^1 \leftarrow f_\text{AV-TSE}(y, v; \theta)$
    \EndIf
    
    \State \textbf{Stage 2: Memory-Enhanced Extraction}
    \State Compute curriculum weight: $\alpha \leftarrow \min(1, \frac{ep}{ep_\text{cr}})$
    \State Apply curriculum learning: $\hat{x}^1_\text{cr} \leftarrow \alpha \cdot \hat{x}^1 + (1-\alpha) \cdot \frac{\|\hat{x}^1\|^2}{\|x\|^2} \cdot x$
    
    \State Initialize memory bank: $\mathcal{M} \leftarrow \emptyset$
    \For{$i = 1$ to $N$}
        \State $\hat{x}^1_{\text{cut}} \leftarrow \hat{x}^1_{\text{cr}[0:T-i \cdot T_{\text{sh}}]}$
        \State $\hat{x}^1_{\text{pad}}  \leftarrow \text{LeftZeroPad}(\hat{x}^1_{\text{cut}}, T)$
        \State $\mathcal{M} \leftarrow \mathcal{M} \cup \{\hat{x}^1_{\text{pad}}\}$
    \EndFor
    
    \State $\mathcal{M}_{\text{shuf}} \leftarrow \text{RandomShuffle}(\mathcal{M})$
    \State $\hat{x}^2 \leftarrow f_\text{AV-TSE}(y, v, \mathcal{M}_{\text{shuf}}; \theta)$
    
    \State \textbf{Loss Computation}
    \State $\mathcal{L} \leftarrow \beta \mathcal{L}_{\text{SI-SNR}}(\hat{x}^1, x) + (1 - \beta) \mathcal{L}_{\text{SI-SNR}}(\hat{x}^2, x)$
    \State \textbf{return} $\theta^*$ after optimization
    \end{algorithmic}
\end{algorithm}

\vspace{-2mm}
\subsubsection{PAR Training Strategy} 

As shown in Algorithm \ref{alg}, the proposed PAR  strategy contains two stages. 
The first stage aims to get a self-enrolled estimated speech $\hat{x}^1$. The second stage utilizes  $\hat{x}^1$ as a new reference to get the final output $\hat{x}^2$. 

\textbf{Stage 1:} The model performs extraction using only the visual cue $v$ as the  reference. While for speaker bank training, we find that utilizing an additional pre-enrolled speech $p$ which is easy to obtain in real-world scenarios, contributes to improving the quality of  $\hat{x}^1$.  We refer to this initial setting as \textit{VP$_\text{Init}$} (applied only to the speaker bank). The alternative initialization, where only the visual cue $v$ is used without the pre-enrolled speech, is referred to as  \textit{V$_\text{Init}$} (applied to both the speaker bank and contextual bank).


\textbf{Stage 2:} We utilize the self-enrolled estimated speech $\hat{x}^1$ as reference together with visual cues. However, at early training epochs, the quality of $\hat{x}^1$ is often poor because the model has not yet converged, which can destabilize training  \cite{li2019listening,li2018source,wang2020online,andreev23_interspeech,pan24_interspeech}. To mitigate this issue, we adopt a curriculum learning strategy \cite{bengio2009curriculum, williams1989learning, bengio2015scheduled}, gradually transitioning from the ground truth target speech $x$ to the self-enrolled speech $\hat{x}^1$. Specifically, we define a new curriculum-guided speech  $\hat{x}^1_\text{cr}$, where the curriculum weight $\alpha$ increases linearly from 0 to 1 with the epoch index $ep$ over the course of the first $ep_\text{cr}$ epochs. After that, $\hat{x}^1_\text{cr}$ fully degenerates into $\hat{x}^1$, enabling the model to rely solely on its own self-enrollment speech. Cause we use  signal-to-noise ratio (SI-SNR) as the training loss, the energy of $\hat{x}^1$ is not stable. Hence $\frac{\|\hat{x}^1\|^2}{\|x\|^2}$ is used to reduce the energy gap. 
To teach the model how to select the most relevant slot from memory bank, we create a memory bank $\mathcal{M}$ to mimic this processing. $\hat{x}^1_\text{cr}$ is  segmented into $N$ overlapping speech, with each shifts $T_\text{sh}$ samples. Each segment is left-padded to the original  length $T$ to maintain shape consistency. To avoid overfitting to temporal continuity, the memory bank is randomly shuffled before being used as the reference in  training. 
The training loss will be computed both on $\hat{x}^1$ and  $\hat{x}^2$,  and could be expressed as:
\vspace{-1mm}
 \begin{equation}
\mathcal{L}= \beta  \mathcal{L}_{\text{SI-SNR}}( \hat{x}^1,x) + \left(1-\beta \right) \mathcal{L}_{\text{SI-SNR}}( \hat{x}^2,x),
 \label{equ:loss}
 \vspace{-1mm}
 \end{equation}
 where $\mathcal{L}_{\text{SI-SNR}}$ is defined as: 
 \vspace{-1mm}
 \begin{equation}
    \mathcal{L}_{\text{SI-SNR}}(\hat{x},x) = -10 \log_{10}\frac{||\frac{<\hat{x},x>x}{||x||^2}||^2}{||\hat{x}-\frac{<\hat{x},x>x}{||x||^2}||}.
    \vspace{-1mm}
\end{equation}

\subsubsection{Online Inference Strategy}
\label{sec:online_infer}

\begin{figure*}[htpb]
    \centering
    \includegraphics[width=0.98\linewidth]{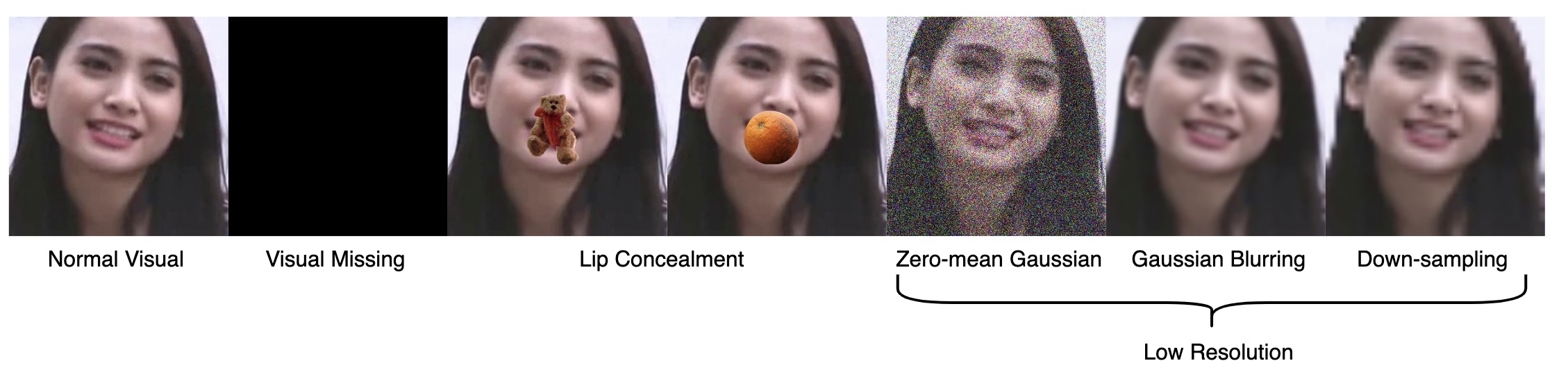}
    \vspace{-3mm}
    \caption{Examples of visual impairments.}
    \label{fig:data}
    \vspace{-3mm}
\end{figure*}

In online real-time processing \cite{li2024momuse,pan2024neuroheed,pan2024neuroheed+}, the model begins processing once the accumulated input reaches a predefined initialization length, denoted as $T_\text{init}$, which corresponds to the $0$-th window step. The first stage in Algorithm~\ref{alg} simulates this initialization step. After that, the model processes speech using a sliding window of length $T_\text{win}$, with the window shifting forward by $T_\text{sh}$ at each step.
All parameters are summarized in Table \ref{tab:parameters}. 

\vspace{-1mm}
\begin{table}[ht]
    \centering
    \caption{Online processing parameters descriptions.}
    \vspace{-1mm}
    \begin{tabular}{@{}ll@{}}
        \toprule
        \textbf{Parameter} & \textbf{Description} \\ \midrule
        \multirow{2}{*}{$T_\text{win}$}& Duration of sliding window (in samples). \\
        & $T_\text{win}$ is the receptive field of models. \\ \midrule
        \multirow{2}{*}{$T_\text{sh}$} & Duration of window shifting length (in samples). \\
        & The sliding window moves $T_\text{sh}$ every time.  \\ \midrule
        \multirow{2}{*}{$T_\text{init}$} & Duration of cold start time (in samples). \\
        & The incoming data accumulates to $T_\text{init}$  before starting.        
        \\\bottomrule
    \end{tabular}
    \label{tab:parameters}
    \vspace{-1mm}
\end{table}
In most cases, the model uses only visual cues as reference during the initialization step. However, in the VP$_\text{Init}$ setting of the speaker bank, both the visual cue and the pre-enrolled speech are used as reference for initialization.

Cause we use $\mathcal{L}_\text{SI-SNR}$ as loss function, the energy of estimated speech for each window step is different. To ensure the perception consistency, we utilize the energy normalization strategy applied in \cite{pan2024neuroheed} to smooth the overall output. We normalize $\hat{x}(k)$ based on the energy between current window step $k$ and the cumulative energy of the outputs from previous steps $0$ through $k-1$: 
\begin{equation}
    \hat{x}(k)= \begin{cases}
        \gamma\cdot\frac{\hat{x}(k)}{\| \hat{x}(k)\|^2} & k=0\\
         \hat{x}(k) \cdot \frac{\|\hat{x}_{0:k-1} \|^2}{\|\hat{x}(k) \|^2}  & \text{otherwise},
    \end{cases}
    \label{equ:norm}
\end{equation}
where $\|\hat{x}_{0:k-1}\|^2$ denotes the cumulative energy of all previous window outputs from step $0$ to $k-1$, and $\gamma$ is a normalization weight aims to control the energy of the speech.

\vspace{-2mm}
\section{Experimental setup}
\subsection{Dataset}

\subsubsection{Visual Impairments}
\label{sec:data}
  We simulate three types of impaired scenarios: visual missing, lip concealment, and low resolution, as shown in Fig. \ref{fig:data}. 
\begin{itemize}
    \item \textbf{Visual missing}: the speaker’s face is undetected, hence the whole visual frame is set to zero values.
    \item \textbf{Lip concealment}: Target speaker's lip is occluded by a certain obstacle, such as hands or microphones. We utilize the  Naturalistic Occlusion Generation (NatOcc) dataset~\cite{Voo2022DelvingIH} to simulate this, similar to ~\cite{Hong2023WatchOL} \footnote{\url{https://github.com/joannahong/AV-RelScore}}. The NatOcc patches consist of various objects, such as fruits, desserts, cups, and so on.
    For evaluation fairness,  we reserve one specific object for the test set, and other objects are randomly selected on-the-fly during training. 
    \item \textbf{Low resolution}: Visual frames are not recorded clearly and appear blurry. We utilize three methods to simulate this: zero-mean Gaussian noise \footnote{\url{https://scikit-image.org/docs/0.23.x/api/skimage.util.html\#skimage.util.random_noise}}, Gaussian blurring \footnote{\url{https://pytorch.org/vision/main/generated/torchvision.transforms.GaussianBlur.html}} and image down-sampling. 
    For the training and validation sets,  zero-mean Gaussian noise and Gaussian blurring are selected with equal probability to capture a range of blurriness levels. For the test set, only image down-sampling is applied. 
\end{itemize}

\subsubsection{VoxCeleb-2mix}
 We construct a two-speaker mixture dataset from the VoxCeleb2 dataset \cite{voxceleb2}, where each mixture consists of a target speaker and an interfering speaker. To ensure diversity of speaker and utterances, we select a wide range subset from VoxCeleb2, including 48,000 utterances from 800 speakers for training set and validation set simulation and 36,237 utterances from 118 speakers for test set simulation, following  \cite{muse,wu2024target_cvpr,ijcnn}. Specifically, we simulate 20,000 utterances for the training set, 5,000 for the validation set, and 3,000 for the test set. Each target utterance is mixed with an interfering utterance at a random signal-to-noise ratio (SNR) between -10 dB and 10 dB, temporally aligned with the short utterance. The audio sampling rate $\text{sr}^a$ is 16,000 Hz, and the video frame rate $\text{sr}^v$ is 25 FPS.  None of the speakers in the test set were encountered during training and validation. Utterances are limited to 4-6 seconds during training and real duration during inference.  For each utterance,  only one type of impairment has been selected. The impairment ratio is randomly chosen from [0\%, 80\%) for the training and validation sets, and from [0\%, 100\%) for the test set.

\vspace{-3mm}
\begin{figure}[htbp]
    \centering
    \includegraphics[width=0.49\textwidth]{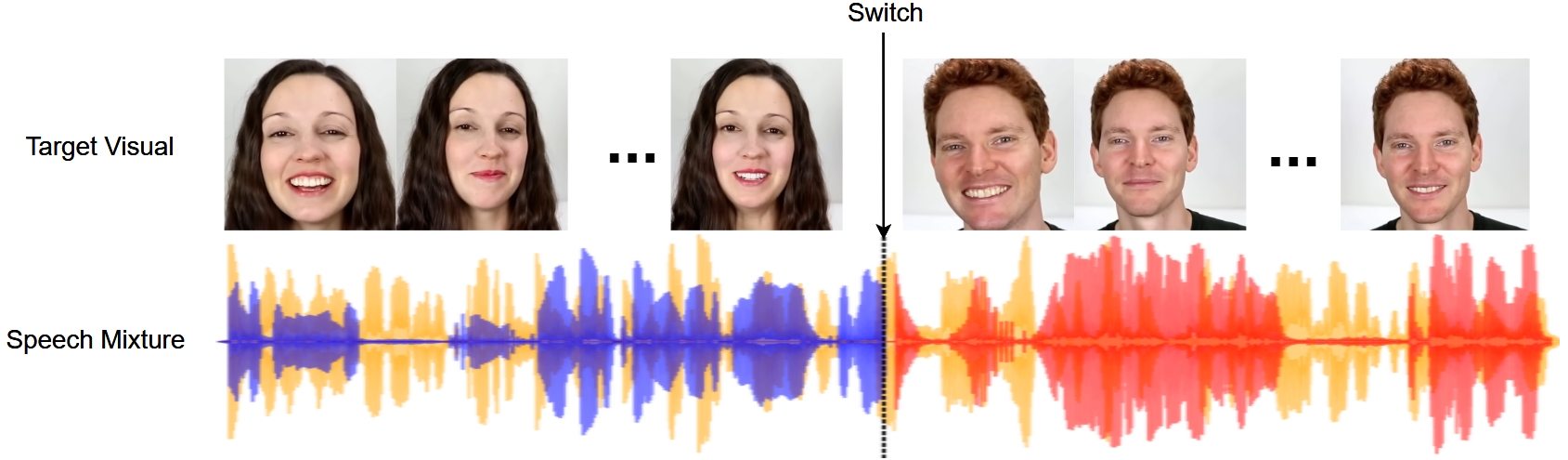}
    \vspace{-5mm}
    \caption{Attention shifts from one target speaker to another at a certain point, while interfering signals persist. }
    \label{fig:switch}
    \vspace{-3mm}
\end{figure}
 \subsubsection{VoxCeleb-2mix-switch}
We build a test set to simulate two target speakers switching in a cocktail party, as shown in Fig. \ref{fig:switch}. To make sure there are always two speakers speaking, the interfering speaker simultaneously speaks for at least 10 seconds. The switch point occurs randomly between 4 and 6 seconds. 
To achieve this, we select utterances within at least 10 seconds from the VoxCeleb2 test set and simulate 3000 mixtures.
The visual cue impaired setting is the same as VoxCeleb-2mix, except we also ensure visual cues are clean for 1 second after the switch point,  for smooth cold start purposes. Other settings are the same as VoxCeleb-2mix.

\vspace{-3mm}
\subsection{Metrics}  
\vspace{-1mm}
We select the SI-SNR \cite{le2019sdr}, SDR  \footnote{\url{https://pypi.org/project/fast-bss-eval/}}\cite{le2019sdr},   perceptual evaluation of speech quality (PESQ) \footnote{\url{https://pypi.org/project/pesq/}} \cite{rix2001perceptual} 
and the short-term objective intelligibility (STOI) \footnote{\url{https://pypi.org/project/pystoi/}}\cite{taal2010short}
as metrics to evaluate signal similarity and speech quality.
 Higher values denote better performance. 
  In online evaluation, we use a real-time factor (RTF) to express
the model’s computational complexity, the lower the better. It is calculated using an NVIDIA Tesla V100 GPU, and is given by the following formula:
\begin{equation}
\text{RTF} = \frac{\text{Processing Time}}{\text{Speech Length}}.
\end{equation}

\vspace{-4mm}
\subsection{Implementation  Details}
\subsubsection{Training}

The initial learning rate was set to 1e-3 using the Adam optimizer. 
The learning rate was halved if the best validation loss (BVL) did not improve within six consecutive epochs, and the training stopped if the BVL did not improve within ten consecutive epochs. The maximum number of training epochs was set to 100.  The $ep_\text{cr}$ was set to 50 for curriculum learning. The number of slots $N$ in memory banks was randomly chosen from $[1, 5]$. During training, the window shift $T_\text{sh}$   was set within $[0, \text{sr}^a]$. The loss weight $\beta$ was set to $0.2$. 

\subsubsection{Inference}
The online parameters $T_\text{win}$, $T_\text{sh}$, and $T_\text{init}$ were set to $2  \text{sr}^a$, $0.2  \text{sr}^a$, and $2 \text{sr}^a$, respectively. 
The audio sampling rate $\text{sr}^a$ is 16,000 Hz. 
In the 0-th window step, clean visual frames without impairments were used to ensure high-quality self-enrolled speech. The number of slots $N$ in each memory bank was set to 1 by default, and the duration of self-enrolled speech was also fixed at 2 seconds by default.  The normalization weight $\gamma$ was set to 0.7. 

\subsubsection{Dataset}
The visual inputs were cut from the middle part of all face frames and transformed into grayscale with the size of $112 \times 112$. 
 We applied visual impairments directly on video frames. For lip concealment case, the obstacle object was randomly positioned around the center of each frame with offsets of $\left[13, 17 \right]$. 
For zero-mean Gaussian, the noise variance was randomly selected from $[0.02, 0.2]$. For Gaussian blurring,  a 2D Gaussian filter with a kernel size of  $(13, 13)$ and standard deviations of $(4, 8)$ was applied. For down-sampling, the resolution of video frames was reduced by a factor of 10. 

 \begin{table}[htbp]
    \centering
    \caption{Four types of evaluation settings in offline mode and online mode. We take the $k$-th window step as an example in online mode. Settings in gray represent the use of static reference inputs (i.e., pre-enrolled or ground-truth target speech) for comparison purposes. The underlined settings are the primary and novel settings proposed in this paper.}
    \begin{tabular}{c|c|c}
    \midrule
         \textbf{Evaluation}& \textbf{Evalation}  & \multirow{2}{*}{\textbf{Formulation}} \\ 
         \textbf{Settings}  & \textbf{mode} & \\ \midrule 
         \multirow{2}{*}{Visual Only}&Offline &$\hat{x} = f_\text{AV-TSE}(y,v ;\theta)$ \\ 
         & Online & $\hat{x}(k) = f_\text{AV-TSE}(y(k),v(k) ;\theta)$ \\ \midrule
         \multirow{2}{*}{\underline{Visual+SelfEnro}} & Offline &  $\hat{x} = f_\text{AV-TSE}(y,v,f_\text{AV-TSE}(y,v;\theta);\theta)$ \\
         & \underline{Online} & $\hat{x}(k) = f_\text{AV-TSE}(y(k),v(k), \hat{x}(k-1);\theta)$ \\ \midrule
                 \rowcolor{gray!20}  \multirow{2}{*}{Visual+PreEnro} &Online & $\hat{x} = f_\text{AV-TSE}(y,v, p;\theta)$ \\
        \rowcolor{gray!20} \multirow{-2}{*}{Visual+PreEnro} & Online & $\hat{x}(k) = f_\text{AV-TSE}(y(k),v(k), p;\theta)$ \\ \midrule 
         \rowcolor{gray!20} \multirow{2}{*}{Visual+TgtEnro} &Offline & $\hat{x} = f_\text{AV-TSE}(y,v, x;\theta)$ \\
         \rowcolor{gray!20}  \multirow{-2}{*}{Visual+TgtEnro}& Online & $\hat{x}(k) = f_\text{AV-TSE}(y(k),v(k), x(k);\theta)$ \\ \midrule
    \end{tabular}
    \label{tab:setting}
    \vspace{-3mm}
\end{table}

\begin{table*}[htbp]
\caption{Comparison between  different speaker banks  in offline and online evaluation mode in terms of SI-SNR. $\text{V}_\text{Init}$ and $\text{VP}_\text{Init}$ are different initial settings for speaker bank, as defined  in Algorithm \ref{alg}.  N/A indicates that the item is not applicable in the given setting. Number of slots $N$ in online mode is set to 1 by default during the inference stage.}
\begin{tabular}{c|c|c|c|c|c|c|c|c|c|c|c|c}
\midrule
\multirow{4}{*}{Sys. \#}& \multirow{4}{*}{Model}&   \multicolumn{2}{c|}{\multirow{2}{*}{Speaker}} &  \multirow{4}{*}{Evaluation}  & \multicolumn{8}{c}{SI-SNR (dB) $\uparrow$} \\  \cmidrule{6-13}

& & \multicolumn{2}{c|}{Bank}& & \multicolumn{4}{c|}{Clean Visual } & \multicolumn{4}{c}{Impaired Visual } \\ \cmidrule{3-4}   \cmidrule{6-13}

& & \multirow{2}{*}{$\text{V}_\text{Init}$} & \multirow{2}{*}{$\text{VP}_\text{Init}$} &  \multirow{2}{*}{Mode}& Visual & \underline{Visual+}&\cellcolor{gray!20}Visual+ &  \cellcolor{gray!20} Visual+&  Visual & \underline{Visual+} &\cellcolor{gray!20}Visual+ &  \cellcolor{gray!20}Visual+  \\

&  & & & & Only& \underline{SelfEnro} & \cellcolor{gray!20}PreEnro & \cellcolor{gray!20} TgtEnro & Only & \underline{SelfEnro} &\cellcolor{gray!20} PreEnro &   \cellcolor{gray!20}TgtEnro \\ \midrule 

& Mixture   & & & & \multicolumn{8}{c}{0 } \\ \midrule 
1& TDSE \cite{tdse}  &  & & \multirow{3}{*}{Offline}   &11.16 & \multicolumn{3}{c|}{N/A}  & 10.32 &\multicolumn{3}{c}{N/A}  \\ \cmidrule{1-4} 
2&\multirow{2}{*}{MeMo (TDSE)}  &  \ding{52}  & & & 11.23 &  11.31  & \cellcolor{gray!20}11.31 &  \cellcolor{gray!20}11.40 &10.25 &10.35  &\cellcolor{gray!20} 10.52 &  \cellcolor{gray!20}10.69 \\ \cmidrule{1-1}
3&  & & \ding{52}&  & N/A&  \textbf{11.46} & \cellcolor{gray!20}11.33  &  \cellcolor{gray!20}11.61 &N/A & \textbf{10.84}  & \cellcolor{gray!20}10.83&   \cellcolor{gray!20}11.27 \\ \hline
1& TDSE \cite{tdse}  &  & & \multirow{3}{*}{Online}   &9.55 & \multicolumn{3}{c|}{N/A}  & 8.13&\multicolumn{3}{c}{N/A}  \\ \cmidrule{1-4} 

2&\multirow{2}{*}{MeMo (TDSE)}  &  \ding{52}  & & & 9.65 &  9.87 & \cellcolor{gray!20}9.91  &  \cellcolor{gray!20}10.07 &8.16 &8.64  & \cellcolor{gray!20}8.60 &  \cellcolor{gray!20}8.76 \\ \cmidrule{1-1}
3&  & & \ding{52}&  & N/A&  \textbf{10.12}  & \cellcolor{gray!20}10.05 &  \cellcolor{gray!20}10.48 &N/A & \textbf{9.43}  &\cellcolor{gray!20}9.24 &   \cellcolor{gray!20}9.85 \\ \hline
\end{tabular} 
\vspace{-3mm}
\label{tab:speaker bank}
\end{table*}

\begin{table}[htbp]
\centering
\caption{The impact of varying the number of memory slots in the online mode under the `Visual+SelfEnro' setting for the impaired visual scenario. The length of self-enrolled speech refers to the audio duration used for self-enrollment, before being embedded into speaker embeddings. All results are reported in terms of SI-SNR (dB). }
\begin{tabular}{c|c|c|c|c|c}
\midrule
\multirow{4}{*}{Sys. \#}  & Length& \multicolumn{4}{c}{Number of Slots $N$} \\ \cmidrule{3-6}

 & of & \multirow{3}{*}{1} & \multirow{3}{*}{2} & \multicolumn{2}{c}{4} \\ \cmidrule{5-6}

& Self-enrolled &   & & \multicolumn{2}{c}{Popping Methods} \\ \cmidrule{5-6}
& Speech (seconds) &  & & FIFO & ABS \\ \midrule

\multirow{3}{*}{2}  & 2s & 8.64 & 8.66 & 8.66 & \textbf{8.69} \\ 
& 1s & 8.62 & 8.64&  8.64&8.67\\ 
& 0.5s &8.60 & 8.60 &8.62&8.67\\ \midrule
 \multirow{3}{*}{3}& 2s & 9.43 & 9.43 & 9.44 &\textbf{9.48} \\ 
&  1s & 9.26 &9.31 &9.31 &9.48 \\
&  0.5s & 8.81 &8.99 &9.15 &9.30 \\ \midrule
\end{tabular} 
\label{tab:slot_SB}
\vspace{-3mm}
\end{table}

\begin{table*}[htbp]
\centering
\caption{
Performance of contextual bank under different evaluation settings. Here we only present the speaker bank with $\text{VP}_\text{Init}$. The number of slots $N$ in online mode is set to 1 by default. }
\begin{tabular}{c|c|c|c|c|c|c|c|c|c|c|c|c}
\midrule
\multirow{4}{*}{Sys. \#}& \multirow{4}{*}{Model}&  \multirow{4}{*}{Contextual} &  \multirow{4}{*}{Speaker}& \multirow{4}{*}{Evaluation} & \multicolumn{7}{c}{SI-SNR (dB) $\uparrow$} \\  \cmidrule{6-13}

& & &  &  & \multicolumn{4}{c|}{Clean Visual } & \multicolumn{4}{c}{Impaired Visual } \\   \cmidrule{6-13}

& &  \multirow{2}{*}{Bank}& \multirow{2}{*}{Bank} & \multirow{2}{*}{Mode}& Visual & \underline{Visual+}& \cellcolor{gray!20}Visual+ &  \cellcolor{gray!20}Visual+&  Visual & \underline{Visual+} & \cellcolor{gray!20}Visual+ &  \cellcolor{gray!20}Visual+  \\

&  & & &  & Only& \underline{SelfEnro} &  \cellcolor{gray!20}PreEnro &    \cellcolor{gray!20}TgtEnro & Only & \underline{SelfEnro} &  \cellcolor{gray!20}PreEnro &   \cellcolor{gray!20}TgtEnro\\ \midrule 

& Mixture   && & & \multicolumn{7}{c}{0 } \\ \midrule 
1& TDSE \cite{tdse}  &    &  &\multirow{5}{*}{Offline} &  11.16 & \multicolumn{3}{c|}{N/A}  & 10.32 &\multicolumn{3}{c}{N/A}  \\  \cmidrule{1-4} 
3& \multirow{3}{*}{MeMo (TDSE)} & &  \ding{52} & & N/A&  11.46 &  \cellcolor{gray!20}11.33 &  \cellcolor{gray!20}11.61  &N/A & \textbf{10.84} &   \cellcolor{gray!20}10.83  &  \cellcolor{gray!20}11.27  \\ \cmidrule{1-1}
4&  &   \ding{52}   & & & 11.51&  \textbf{11.63} &    \cellcolor{gray!20}11.01 & \cellcolor{gray!20} 12.37 &  10.59 &  10.70 &  \cellcolor{gray!20}9.54 &  \cellcolor{gray!20} 12.37 \\ \cmidrule{1-1}
5&  & \ding{52} &  \ding{52} & &  N/A&  11.35 &   \cellcolor{gray!20}11.25 &   \cellcolor{gray!20}11.65 &N/A & 10.73 & \cellcolor{gray!20}10.60 & \cellcolor{gray!20} 11.53  \\ \midrule

1& TDSE \cite{tdse}  &    &  &\multirow{5}{*}{Online} &9.55 & \multicolumn{3}{c|}{N/A}  & 8.13 &\multicolumn{3}{c}{N/A}  \\  \cmidrule{1-4} 
3& \multirow{3}{*}{MeMo (TDSE)} & & \ding{52}&  & N/A&  10.12  &  \cellcolor{gray!20}10.05 &  \cellcolor{gray!20}10.48 &N/A & 9.43  &  \cellcolor{gray!20}9.24 &   \cellcolor{gray!20} 9.85 \\ \cmidrule{1-1} 
4&  &   \ding{52}   & &  &  10.24 &   \textbf{10.50}& \cellcolor{gray!20}9.27 &  \cellcolor{gray!20} 10.94&  8.82& \textbf{10.34} & \cellcolor{gray!20}7.71 &  \cellcolor{gray!20}10.93 \\ \cmidrule{1-1}
5&  & \ding{52} &  \ding{52} & &  N/A&  10.33 &   \cellcolor{gray!20}10.64 &   \cellcolor{gray!20}10.01 &N/A  &10.04 &  \cellcolor{gray!20}9.05 &  \cellcolor{gray!20}10.41  \\ \midrule

\end{tabular} 
\vspace{-3mm}
\label{tab:contextual bank}
\end{table*}
\vspace{-2mm}
\section{Results and analysis}

In this section, we evaluate MeMo’s effectiveness across various scenarios and settings, focusing primarily on \textbf{online mode} with \textbf{impaired visuals}. To assess the impact of impaired visuals, we consider two scenarios:
\begin{itemize}
    \item Clean Visual: The visual cues remain intact throughout processing.
    \item Impaired Visual: Some visual frames are partially corrupted as described in Section \ref{sec:data}.
\end{itemize}
For each visual scenario, we evaluate the models under four different settings, as shown in Table~\ref{tab:setting}. These evaluation settings are designed to compare the effectiveness of different types of reference information. Typical AV-TSE systems operate under the `Visual Only' setting, where only visual cues are used. In contrast, our proposed model, MeMo, can leverage dynamic reference, self-enrolled speech from the previous window step under the `\underline{Visual+SelfEnro}' setting,  which is the primary and novel setting introduced in this work.

The `Visual+PreEnro' and `Visual+TgtEnro' settings are similar to `Visual+SelfEnro', but they only use static reference, pre-enrolled speech and ground-truth target speech, respectively. Hence,  memory banks are not applicable under these two settings, since the reference speech does not evolve over time. These settings are included as comparison to help evaluate the relative effectiveness of self-enrolled cues.

\vspace{-2mm}
\subsection{Speaker Bank}
There are two types of initial settings for speaker bank, as shown in Algorithm \ref{alg}. V$_\text{Init}$ represents the case where only visual cues are used for initialization, while VP$_\text{Init}$ incorporates both visual cues and pre-enrolled speech are used for initialization. Table~\ref{tab:speaker bank} presents the overall results of applying the speaker bank. We use TDSE \cite{tdse}, which is a time-domain model based on Conv-TasNet \cite{Conv-TasNet}, as baseline model. In this table, MeMo adopts the same backbone structure like TDSE, but incorporates an additional speaker bank. Several key observations could be drawn from the results:

\begin{itemize}
    \item The performance gap between clean and impaired visual conditions is obvious. This supports our motivation that the quality of visual cues significantly affects extraction performance. 
    \item Compared to offline mode, system performance in online mode is generally lower. This is reasonable since the receptive field in online models is shorter, limiting the available information for extraction.
    \item  MeMo with speaker bank under `Visual+SelfEnro' evaluation setting provides consistent performance improvements than System 1 under `Visual Only' setting, especially in online mode with impaired visual inputs. This demonstrating the superior effectiveness of self-enrollment and attentional momentum. 
    \item Between two initialization methods,  $\text{VP}_\text{Init}$ performs better than $\text{V}_\text{Init}$, especially in online mode with impaired visual cues,  highlighting the benefit of incorporating multiple cues as reference information during initialization.
    \item Among the four evaluation settings, `Visual+TgtEnro' serves as an idealized upper bound.  The `Visual+SelfEnro' setting yields results similar to `Visual+PreEnro' (sometimes `Visual+SelfEnro'  is better) and outperforms the `Visual Only' setting which performs the worst. This indicates that MeMo effectively leverages speaker information from self-enrolled speech.
\end{itemize}

Next, we investigate how the number of slots affects model performance. As shown in Table~\ref{tab:slot_SB}, increasing the number of slots leads to performance gains, particularly for System 3 with $\text{VP}_\text{Init}$. Notably, when the duration of self-enrolled speech is short, the model benefits more from having additional slots in the speaker bank. This suggests that the model can retrieve complementary speaker identity information across different slots. However, when sufficient speaker information is already available, adding more slots yields no significant improvement.
Regarding different popping strategies, the attention-based selection (ABS) method provides a performance boost, though the improvement is relatively modest.

\vspace{-2mm}
\subsection{Contextual Bank}
\label{sec:contextul_bank}
Although the speaker bank contributes to performance improvement, the gains are relatively limited. This observation motivates us to explore the effectiveness of the contextual bank, which is designed to capture temporal and contextual information. In this section, we present a detailed comparison between the speaker bank with VP$_\text{Init}$  and the contextual bank, as shown in Table~\ref{tab:contextual bank}.
We summarize several key observations:

\begin{itemize}
\item 
While the `Visual+SelfEnro' performance of contextual bank and speaker bank is similar in the offline mode, the contextual bank demonstrates obviously better effectiveness in the online mode.

\item The contextual bank yields substantial performance gains over the baseline, particularly under impaired visual conditions in the online mode, which is our primary target scenario, with improvements exceeding 2 dB in SI-SNR.

\item The `Visual+SelfEnro' setting presents a significant improvement over `Visual+PreEnro’ and achieves results in online mode comparable to our upper-bound `Visual+TgtEnro'. This indicates that leveraging online contextual information plays a critical role in enhancing model performance, and our proposed method and training strategy effectively utilize nearly full contextual information.

\item Combining the speaker bank with the contextual bank does not lead to complementary improvements. On the contrary, it degrades performance, resulting in worse outcomes than using the contextual bank alone, possibly due to redundancy or conflicting reference information.
\end{itemize}

\begin{table}[htbp]
\centering
\caption{The impact of varying the number of memory slots in the online mode under the `Visual+SelfEnro' setting for the impaired visual scenario. All results are reported in  SI-SNR (dB). }
\begin{tabular}{c|c|c|c|c|c}
\midrule
\multirow{4}{*}{Sys. \#}  & Length& \multicolumn{4}{c}{Number of Slots $N$} \\ \cmidrule{3-6}

 & of & \multirow{3}{*}{1} & \multirow{3}{*}{2} & \multicolumn{2}{c}{4} \\ \cmidrule{5-6}

& Self-enrolled &   & & \multicolumn{2}{c}{Popping Methods} \\ \cmidrule{5-6}
& Speech (seconds) &  & & FIFO & ABS \\ \midrule

4  & 2s & \textbf{10.34} & 10.10 & 9.33 & 9.11 \\ \midrule
5&  2s &  \textbf{10.04} & 9.83 &9.62 &9.54 \\\midrule 
\end{tabular} 
\label{tab:slot_CB}
\vspace{-3mm}
\end{table}

\begin{table}[htbp]
\centering
\caption{Performance of the contextual bank using contextual embeddings from different time steps under impaired visuals with `Visual+SelfEnro' setting. The number of Slots $N$ is set to 1. }
\begin{tabular}{c|c|c}
    \midrule
    System & SI-SNR (dB) & Remark \\ \hline
    \multirow{4}{*}{System~4} 
    & 10.34 & Latest contextual embedding \\ 
    & 9.66  & Second-latest contextual embedding \\  
    & 8.91  & Third-latest contextual embedding \\ 
    & 8.42  & Fourth-latest contextual embedding \\ \hline
\end{tabular}
\label{tab:contextual_step1}
\vspace{-3mm}
\end{table}

Similarly, we analyze the impact of the number of slots in the contextual bank, as shown in Table~\ref{tab:slot_CB}. As the number of slots increases, performance gradually degrades. This aligns with our intuition: in the case of the contextual bank, the latest contextual information is typically more relevant for current speaker extraction, and adding excessive historical information may introduce noise or conflicts. Regarding different popping strategies, the attention-based selection (ABS) method does not improve the performance of the contextual bank. This may be because contextual speech embeddings are relatively low-level representations, making it difficult for the attention mechanism to assign correct scores to different slots. 
To further investigate this issue, we evaluate System~4 using contextual embeddings extracted from different time steps. As shown in Table~\ref{tab:contextual_step1}, embeddings from the most recent time step yield the largest performance gains, indicating that contextual information is highly time-sensitive. This observation suggests that the latest contextual embedding carries the most informative and relevant cues for the task.
In addition, when applying ABS to the Contextual Bank, the attention mechanism does not consistently assign the highest weight to the latest contextual embedding. Specifically, we observed that in approximately 95\% of the cases, the latest contextual embedding receives the highest attention score. This imperfect alignment leads to suboptimal retrieval, which helps explain why ABS performs slightly worse than the simpler FIFO strategy when applied to contextual features.

\begin{table}[htbp]
    \centering
        \caption{Comparison results across different types of impaired visual cues under the online mode. The number of slots $N$ for each system is set to 1 by default in this table. Results are reported in  SI-SNR (dB).}
    \begin{tabular}{c|c|c|c|c|c}
    \midrule
         \multirow{3}{*}{Sys. \#} & \multicolumn{4}{c|}{Impaired }  & Clean \\ \cmidrule{2-6}
         & Visual & Lip & Low & \multirow{2}{*}{Avg.} &\multirow{2}{*}{Avg.}  \\
         & Missing & Concealment & Resolution &  & \\ \hline
         1 & 6.35  & 9.05 &8.98 & 8.13 & 9.55 \\ 
         2 & 7.28 & 9.59 & 9. 03 & 8.64 & 9.65  \\
         3 & 8.57 & 9.96 & 9.75 & 9.43  & 10.12\\ 
         4 & \textbf{10.18} & 10.31 & \textbf{10.52} & \textbf{10.34} & \textbf{10.50}  \\
         5 & 9.61 & \textbf{10.46} & 10.05 & 10.04 & 10.33\\ \hline
    \end{tabular}
    \label{tab:type}
\end{table}

\begin{table}[htbp]
\centering
\caption{Comparison of whether using clean visual cues in the $0$-th window step for self-enrollment speech (initialization step).  These results are reported in SI-SNR (dB). }
\begin{tabular}{c| c |c |c |c}
\midrule
\diagbox{Visual Init}{\\Sys. \#} &   2 & 3 & 4 & 5 \\  \midrule 

Clean Visual &    8.64 & 9.43 &  10.34 & 10.04  \\ \midrule
Impaired Visual & 7.33 & 8.53 &8.91 &  9.11 \\ \midrule
\end{tabular} 
\label{tab:init}
\vspace{-3mm}
\end{table}

\begin{table*}[hbp]
\centering
\caption{Comparison of SI-SNR and RTF for System 4 with contextual bank across different window length $T_\text{win}$. $T_\text{init}$ and $T_\text{sh}$ are set to $2$s and $0.2$s, respectively. }
\begin{tabular}{c|c| c |c |c |c |c|c|c|c|c}
\midrule
\multirow{2}{*}{Sys. \#} & \multicolumn{2}{c|}{1s} & \multicolumn{2}{c|}{1.5s} & \multicolumn{2}{c|}{2s} & \multicolumn{2}{c|}{3s} &\multicolumn{2}{c}{4s} \\ 
& SI-SNR (dB) $\uparrow$ & RTF $\downarrow$& SI-SNR (dB) $\uparrow$ & RTF $\downarrow$& SI-SNR (dB) $\uparrow$ & RTF $\downarrow$& SI-SNR (dB) $\uparrow$ & RTF $\downarrow$& SI-SNR (dB) $\uparrow$ & RTF $\downarrow$ \\ \midrule
4 & 8.55 &0.1082 & 9.86& 0.1093 &10.34& 0.1103 & 10.40& 0.1141 & 10.45& 0.1422 \\ \midrule
\end{tabular} 
\label{tab:win}
\end{table*}
\begin{table*}[hbp]
\centering
\caption{Comparison of SI-SNR and RTF for System 4 with contextual bank across different shift length $T_\text{sh}$. $T_\text{init}$ and $T_\text{win}$ are set to $2$s and $2$s, respectively. }
\begin{tabular}{c|c| c |c |c |c |c|c|c}
\midrule
\multirow{2}{*}{Sys. \#} & \multicolumn{2}{c|}{0.05s} & \multicolumn{2}{c|}{0.1s} & \multicolumn{2}{c|}{0.2s} & \multicolumn{2}{c}{0.3s}  \\ 
& SI-SNR (dB) $\uparrow$ & RTF $\downarrow$& SI-SNR (dB) $\uparrow$ & RTF $\downarrow$& SI-SNR  (dB) $\uparrow$ & RTF $\downarrow$ & SI-SNR (dB) $\uparrow$ & RTF $\downarrow$ \\ \midrule
4 & 8.84& 0.5233 & 9.01 & 0.2341& 10.34 & 0.1103& 7.78& 0.0733 \\ \midrule
\end{tabular} 
\label{tab:sh}
\vspace{-3mm}
\end{table*}

\subsection{Impact of the Impairment Types}
\label{sec:type}
As shown in Table \ref{tab:type}, we report performance across three different types of impairments. For System 1, we evaluate its performance under the `Visual Only' setting, while for other systems, we report their performances under the `Visual+SelfEnro' setting. 

All impairments negatively impact model performance. Among them, missing visuals have the most severe effect, while the other two types of impairments exhibit similar levels of degradation. However, the 
system with the contextual bank (System 4) maintains similar performance across different impairments and shows minimal degradation, indicating that contextual information can effectively compensate for degraded visuals. 

In contrast, systems with speaker bank suffer significant degradation, especially when visuals are completely missing, although they still outperform System 1. This indicates that these systems remain heavily dependent on visual information, and the inclusion of additional speaker cues alone is insufficient to sustain performance.

\subsection{Impact of Visual Initialization}
As shown in Table \ref{tab:init}, if we use impaired visual sequences as initialization in the $0$-th window step, all  systems' performance drops. This indicates that the quality of self-enrollment speech matters. If the self-enrollment contains errors, these errors could accumulate during the online processing. Therefore, it is better to use high-quality visual cues for initialization. 
\begin{figure}[htbp]
    \centering
    \includegraphics[width=0.5\textwidth]{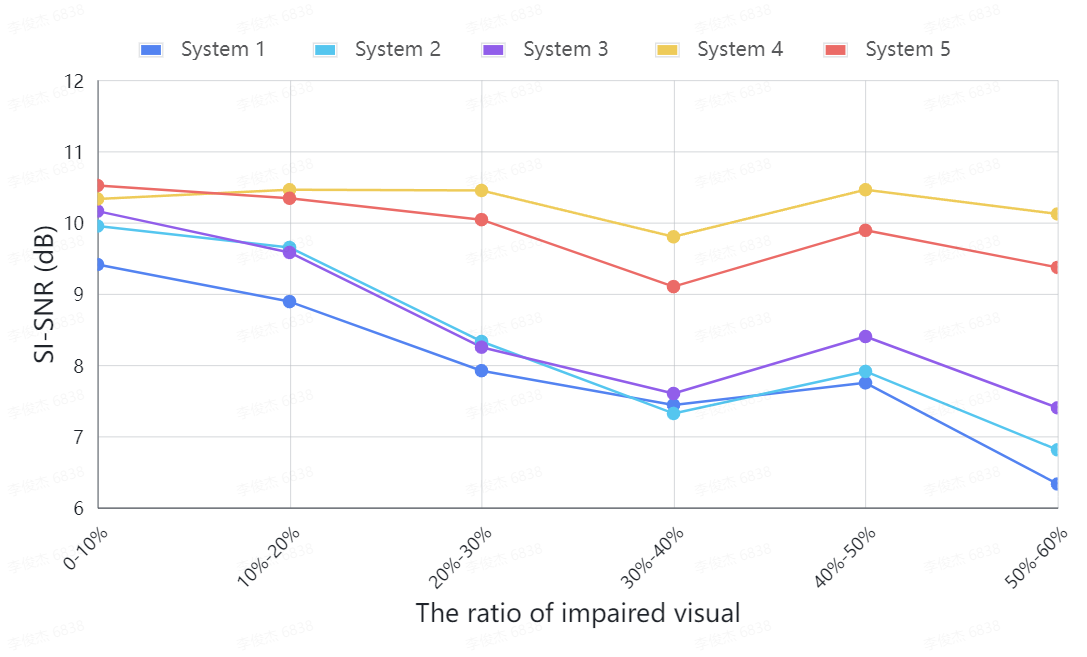}
    \caption{The performance of different systems on different impaired ratios in online mode. 
    Since in $0$-th window step we set visual frames to clean, the impaired ratio of whole utterance is  lessly higher than 60\%.  }
    \label{fig:ratio}
    \vspace{-3mm}
\end{figure}

\vspace{-2mm}
\subsection{Impact of Visual Impaired Ratio}

Fig. \ref{fig:ratio} shows the performance across different impaired ratios of visual. We report System 1's performance under `Visual Only' and other systems under `Visual+SelfEnro'. 

System 1 progressively deteriorates as the impairment ratio increases, which is expected. Systems 2 and 3 outperform the baseline model but still exhibit a similar trend. In contrast, Systems 4 and 5, which leverage contextual information, maintain nearly constant performance, showing minimal impact from impairments, as observed in Section \ref{sec:type}. This indicates that contextual information can effectively compensate for impaired visuals, providing a robust solution to visual impairments.

\vspace{-2mm}
\subsection{Ablation Studies}
In this section, we examine the impact of various online parameters on both performance, measured by SI-SNR, and computational complexity, assessed via the real-time factor (RTF). The RTF is calculated on a single NVIDIA Tesla V100 GPU.

The online evaluation is conducted using a sliding window approach with a window length of $T_\text{win}$ , shifting forward by 
$T_\text{sh}$ at each step. The units of 
$T_\text{win}$ and  $T_\text{sh}$  are in samples, but for convenience, we use seconds as the unit in this section. We only report System 4's performance (contextual bank), cause it is our best system. 

\subsubsection{Study on Window Length}
As shown in Table \ref{tab:win},  the RTF increases as the length of $T_\text{win}$ increases because the models need to process a longer utterance for each sliding window. However, the performance in terms of SI-SNR increases due to model's extended  receptive field.

\subsubsection{Study on Window Shifting Length}
As shown in Table \ref{tab:sh}, the RTF decreases as
the length of $T_\text{sh}$ increases, since a larger shift reduces the number of sliding windows required for online processing. The performance in terms of SI-SNR initially improves with increasing $T_\text{sh}$, but degrades when $T_\text{sh}$ becomes too large. We hypothesize that a slightly larger shift $T_\text{sh}$ improves perceptual consistency across sliding windows. This is likely because the normalization process in Equation~\ref{equ:norm} benefits from larger $T_\text{sh}$ values. However, when $T_\text{sh}$ becomes too large, the overlap between windows degrades, leading to a loss of contextual information continuity. This reduction in contextual information may negatively impact the model's ability to maintain attentional momentum, thus degrading performance.

\begin{figure}[htbp]
\vspace{-3mm}
    \centering
    \includegraphics[width=0.45\textwidth]{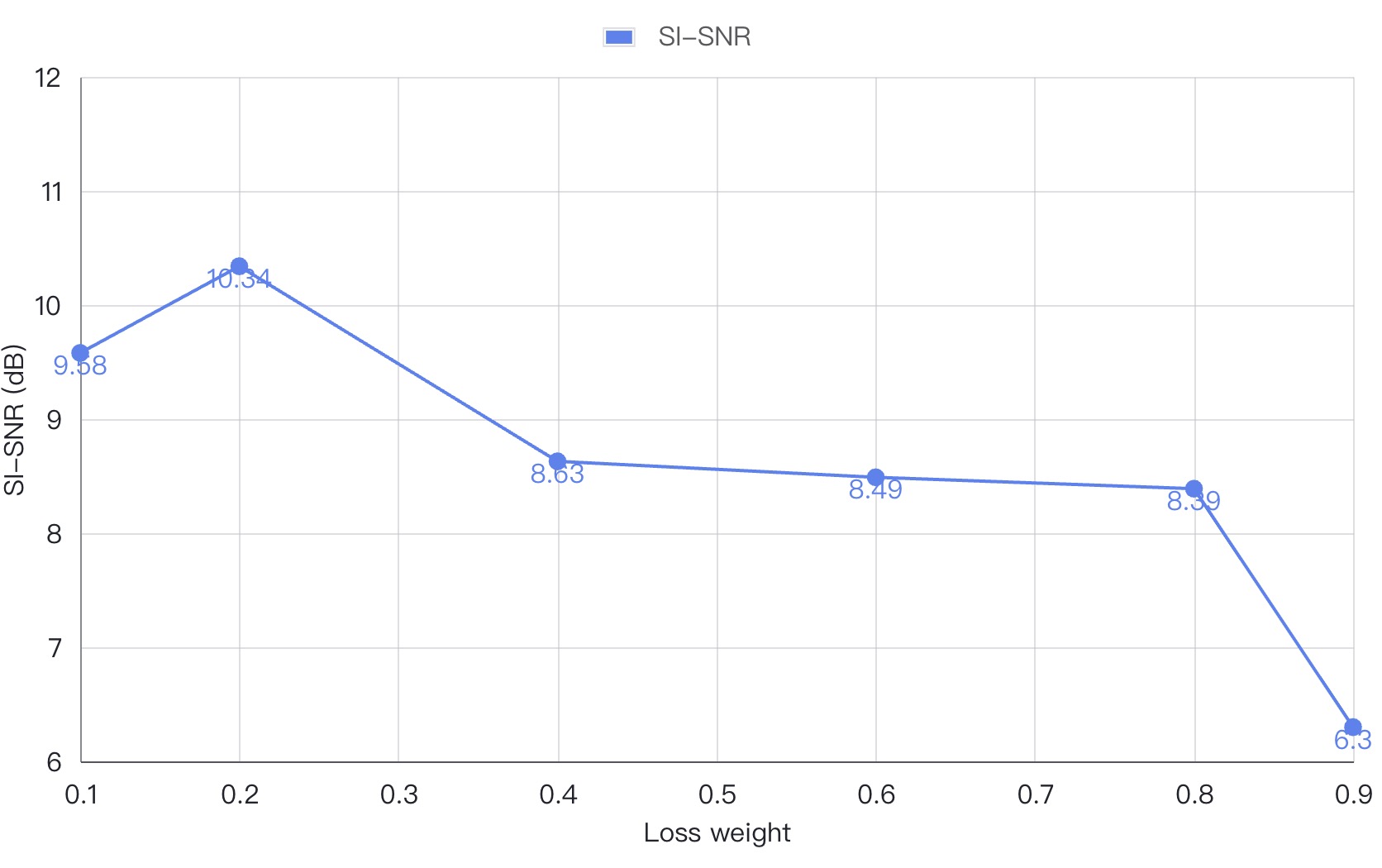}
    \caption{Study for the effect of hyper-parameter $\beta$ in the loss function. }
    \label{fig:weight}
    \vspace{-3mm}
\end{figure}

\subsubsection{Study on Loss Function Weight}
As shown in Fig. \ref{fig:weight}, we compare the effect of different values of the weighting parameter $\beta$ in Equation \ref{equ:loss}. The parameter $\beta$ controls the trade-off between optimization in training Stage 1 and Stage 2. Since our final objective is to improve performance in Stage 2, a relatively small value of $\beta$ is generally preferred. The results in the figure support this conclusion, with the best performance achieved when $\beta = 0.2$.

\subsubsection{Study on Utterance Length}
To evaluate the robustness of different systems under long-duration conditions, we construct an evaluation set by selecting several long utterances from VoxCeleb2 and mixing them to form speech mixtures. Due to the limited availability of long-duration utterances in VoxCeleb2, segmental SI-SNR is reported for utterance lengths up to 40 seconds.

As shown in Table~\ref{tab:long}, System~4 consistently achieves stable performance as the utterance length increases, demonstrating strong robustness in long utterance scenarios. In contrast, the performance of System~1 degrades noticeably with increasing utterance duration. Specifically, its segmental SI-SNR drops from 9.12~dB at 10 seconds to 5.70~dB at 40 seconds.
This performance degradation indicates that relying solely on visual references makes the system more susceptible to accumulated errors and degraded visual cues over time. By incorporating both visual cues and self-enrolled speaker representations, System~4 effectively mitigates this issue and maintains reliable speaker extraction performance over extended durations.

\begin{table}[htbp]
    \centering
     
    \caption{Segmental SI-SNR over  long speech utterance. }
    \begin{tabular}{c|c|c|c|c}
    \hline
        \multirow{2}{*}{System} & \multicolumn{4}{c}{Segmental SI-SNR (dB)} \\ \cline{2-5}
         & 0:10 &0:20 & 0:30 & 0:40  \\ \hline
        1 & 9.12 & 7.24 & 6.38 & 5.70 \\ 
        4 & 11.11 & 10.63 & 10.24 & 10.04 \\ \hline 
    \end{tabular}
    \label{tab:long}
    \vspace{-6mm}
\end{table}

\begin{table*}[htbp]
\centering
\caption{Performance comparison between our proposed methods and other state-of-the-art (SOTA) models under impaired visual scenarios in the online mode. All MeMo systems are configured with the contextual bank only, and their performance is reported under the `Visual+SelfEnro' setting. Star (*) denotes models that are retrained on our impaired datasets for fair comparison.}
\vspace{-1mm}
\begin{tabular}{c|c|c| c |c |c |c}
\midrule
\multirow{3}{*}{Model}  & \multirow{3}{*}{\#Param } &\multirow{3}{*}{ Macs  } & \multicolumn{4}{c}{Impaired Visual} \\ \cmidrule{4-7}
& & & \multirow{2}{*}{SI-SNR (dB) $\uparrow$} &\multirow{2}{*}{SDR (dB) $\uparrow$} &\multirow{2}{*}{PESQ $\uparrow$} & \multirow{2}{*}{STOI $\uparrow$}   \\
&  (M)& (G/s) & & & &   \\ \midrule
Mixture & \multicolumn{2}{c|}{} & 0 & 0.09 & 1.26 & 0.63 \\ \midrule
ImagineNET \cite{ImagineNET}   &  33.61&518.95& 8.97 & 9.40&1.92&0.82 \\ 
MuSE \cite{muse,li2024momuse} &   26.13&21.29 & 8.32 & 8.83 & 1.88 & 0.81  \\
MoMuSE \cite{li2024momuse} &   26.39&21.51 &9.46 & 10.00 & 1.91 & 0.84\\ \cmidrule{1-1} \cmidrule{1-7}

TDSE$^*$ \cite{tdse} &   22.15 & 20.03& 8.13 &8.53 & 1.85 & 0.81  \\
MeMo (TDSE) &   23.00& 20.72 &10.34 &10.76 & 2.08 & 0.84\\ \midrule

USEV$^*$ \cite{usev} & 15.29 & 17.46 & 7.40 &7.76 & 1.69 & 0.79 \\
MeMo (USEV) &   16.14& 18.15 &9.47& 9.85 & 1.91 & 0.83 \\ \midrule


BSRNN$^*$ \cite{luo2023music,wang2024wesep} &  31.17& 224.06 & 7.98& 8.07 & 1.83 & 0.78 \\ 
MeMo (BSRNN) &   32.23 &224.93 & 10.98 & 9.47 & 2.01 & 0.83\\
\midrule

\end{tabular} 
\vspace{-3mm}
\label{tab:models}
\end{table*}

\begin{table}[htbp]
    \centering
    \caption{Performance on the VoxCeleb-2mix-switch test set. In addition to utterance-level SI-SNR, we also report segmental SI-SNR, which evaluates performance over shorter speech segments to better reflect local variations in extraction quality. `Empty' refers to the strategy of emptying the memory bank once a speaker switch point occurs within the model’s receptive field. Note that all models are trained on the VoxCeleb-2mix dataset but  evaluated on the VoxCeleb-2mix-switch dataset.}
    \begin{tabular}{c|c|c|c|c|c|c}
    \hline
        \multirow{2}{*}{Sys.\#} & \multicolumn{5}{c|}{Segmental SI-SNR (dB)} & SI-SNR \\ \cmidrule{2-6}
         & 0:2 &2:4 & 4:8 & 8:12 & 12:16  & (dB) \\ \hline
        1  & 9.07& 7.66 & 5.24 & 5.18 & 7.13  &  5.75\\ \cmidrule{1-1}\cmidrule{2-7}
        4  & 9.26 &8.96  &7.57 & 7.81 & 8.60  &    8.50 \\ \hline
        4+Empty  & 9.26 & 8.97 & 6.99 & 7.76 & 8.38 &   8.06\\ \hline
    \end{tabular}
    \vspace{-3mm}
    \label{tab:switch}
\end{table}

\subsection{Effectiveness on Switching Dataset}
In this section, we investigate whether our proposed method can handle speaker-switching scenarios, as illustrated in Fig.~\ref{fig:switch}. Results are reported in Table \ref{tab:switch}. 
Since MeMo with the contextual bank primarily relies on contextual information, we guess a target speaker switch during a conversation may disrupt the continuity of this context, potentially degrading performance. Hence, we also introduce a `Empty' strategy that manually empties the contextual bank when a speaker switch point is detected within the model's receptive field. 

Overall, System 4 outperforms System 1 in both utterance-level SI-SNR and all segmental SI-SNR evaluations. During the 0–2s interval, only clean visual cues are used as reference; thus, both systems achieve similar performance. After 2s, when the visual cues become impaired, System 1 exhibits a notable performance drop, whereas System 4 degrades only slightly, demonstrating the robustness of leveraging contextual information. In the 4–8s segment, where the speaker switch occurs, both systems experience performance degradation. However, the decline is less pronounced for System 4, demonstrating its resilience under speaker-switching conditions. After the speaker switch, the performance of both systems begins to improve. 

In addition, the `Empty' strategy does not improve performance and even leads to a slight degradation. However, it still performs better than System 1, indicating that our proposed model inherently possesses the ability to handle speaker-switching scenarios with a degree of robustness. This suggests that the contextual bank can effectively adapt to dynamic changes in speaker identity over time. These findings further validate the resilience and adaptability of the proposed MeMo framework in challenging real-time conversational settings.

\vspace{-2mm}
\subsection{Comparison with Other Models}
In this section, we investigate the generalization ability of our proposed framework by applying the best-performing configuration, contextual bank, to various baseline models. As shown in Table~\ref{tab:models}, all models are evaluated under real-time settings. 
Firstly, models equipped with the contextual bank consistently outperform their counterparts across all evaluation metrics. Secondly, the contextual bank introduces only minimal additional parameters and negligible computational complexity.  Finally, compared to prior studies, our proposed framework achieves superior performance on most evaluation metrics.

However, We also evaluate some representative Transformer-based speech separation models, including SEANet \cite{seanet}, IIANet \cite{lee2024iianet}, and Sepformer \cite{subakan2021attention}, within our proposed framework. None of these attention-based models could be optimized stably. We further experimented with replacing global attention with local attention, but this modification did not lead to successful training or performance improvements. 
A plausible explanation lies in the two-stage training protocol adopted in our framework. In the first stage, the model is trained using visual cues only, while in the second stage, the input modality is expanded to include both visual cues and self-enrolled speech. This change in input composition introduces a distribution shift to which Transformer-based architectures appear particularly sensitive in our setting, resulting in degraded optimization stability and generalization. In contrast, CNN- and RNN-based models demonstrate greater robustness to such modality transitions.

 \vspace{-1mm}
\section{Conclusion}
In this paper, we propose MeMo, a general and flexible framework designed to address the challenge of missing visual cues in real-time audio-visual target speaker extraction. 
MeMo achieves attentional momentum through  incorporating two optional adaptive memory banks: the speaker bank and the contextual bank, which leverage historical information to maintain focus on the target speaker over time. While both banks contribute to performance improvements, the contextual bank demonstrates superior effectiveness. Experimental results demonstrate MeMo could yield around 27\% relative improvement compared to the baseline model, which validates the effectiveness of attentional momentum mechanism.

\bibliographystyle{IEEEtran}
\bibliography{refs} 
\end{document}